\documentclass[mathleft
]{an}
\usepackage{graphicx}
\usepackage{times}
\usepackage{natbib}
\bibpunct{(}{)}{;}{a}{}{,}

\begin{document}


\title{Accretion, jets and winds:\\
High-energy emission from young stellar objects\\
Promotionspreis lecture 2010}

\author{H. M. G\"unther\inst{1}\fnmsep\thanks{Corresponding author:
  \email{hguenther@cfa.harvard.edu}\newline}
}
\titlerunning{High-energy emission from young stellar objects}
\authorrunning{H.M. G\"unther}
\institute{Harvard-Smithsonian Center for Astrophysics, 60 Garden Street, Cambridge, MA 02138, USA}

\received{2011}
\accepted{2011}
\publonline{later}


\abstract{%
This article summarizes the processes of high-energy emission in young stellar objects. Stars of spectral type A and B are called Herbig Ae/Be (HAeBe) stars in this stage, all later spectral types are termed classical T Tauri stars (CTTS). Both types are studied by high-resolution X-ray and UV spectroscopy and modeling.
Three mechanisms contribute to the high-energy emission from CTTS: 1) CTTS have active coronae similar to main-sequence stars, 2) the accreted material passes through an accretion shock at the stellar surface, which heats it to a few MK, and 3) some CTTS drive powerful outflows. Shocks within these jets can heat the plasma to X-ray emitting temperatures. Coronae are already well characterized in the literature; for the latter two scenarios models are shown. The magnetic field suppresses motion perpendicular to the field lines in the accretion shock, thus justifying a 1D geometry. The radiative loss is calculated as optically thin emission. A mixture of shocked and coronal gas is fitted to X-ray observations of accreting CTTS. Specifically, the model explains the peculiar line-ratios in the He-like triplets of \ion{Ne}{ix} and \ion{O}{vii}. All stars require only small mass accretion rates to power the X-ray emission.
In contrast, the HAeBe HD 163296 has line ratios similar to coronal sources, indicating that neither a high density nor a strong UV-field is present in the region of the X-ray emission. This could be caused by a shock in its jet. Similar emission is found in the deeply absorbed CTTS DG Tau. Shock velocities between 400 and 500 km~$^{-1}$ are required to explain the observed spectrum. }

\maketitle

\section{Introduction}
Stars and planetary systems form by gravitational collapse of large molecular clouds. Mass infall leads to the formation of a proto-star, which is deeply embedded in an envelope of gas and dust. Due to the conservation of angular momentum matter from this envelope does not accrete directly onto the central star, but forms a proto-stellar disk around it. The envelope depletes and the central star becomes visible as stellar evolution proceeds. In this stage the stars are called classical T~Tauri stars (CTTS), if they are of low-mass ($<3M_{\sun}$, spectral type F or later) and Herbig~Ae/Be stars (HAeBe), if they are of spectral type A or B. CTTS are cool stars with a convective photosphere, thus they generate magnetic fields in solar-like $\alpha-\Omega$ dynamos. Their field lines can reach out a few stellar radii and couple to the disk at the co-rotation radius, because the energetic radiation from the central star ionizes the upper layers of the disk. Thus, the accreting matter is forced to follow the magnetic field lines \citep{1984PASJ...36..105U,1988ApJ...330..350B,1991ApJ...370L..39K,1994ApJ...429..781S}. It is accelerated along the accretion funnel and hits the stellar surface at free fall velocity, so a strong shock develops. This magnetically-funneled accretion scenario explains the wide and unusual emission line profiles observed in CTTS for H$\alpha$ and other Balmer lines \citep{1998ApJ...492..743M,1998AJ....116.2965M}. In fact, CTTS are often defined as young, low-mass stars with an H$\alpha$ equivalent width (EW) $>10$~\AA{}. Once the disk mass drops and the accretion ceases, the width of the H$\alpha$ line decreases and the stars are classified as weak-line T~Tauri stars (WTTS). Eventually, the disk mass is completely absorbed in planets, accreted by the star or driven out of the system by stellar winds and photoevaporation, although transitional disks can exist in WTTS for some time \citep{2006ApJ...645.1283P}.

A wide range of observational evidence supports the magnetically-funneled accretion scenario for CTTS. \citet{2006ApJ...637L.133E} resolved the inner hole in the disk of TW~Hya, the closest CTTS, with radio-interferometry. The accretion funnels are not resolved, but there is a very good agreement between observed and modeled line profiles for H$\alpha$ and other hydrogen emission lines, where emission from infalling material causes the blue-shifted wings of those lines \citep{1998ApJ...492..743M,1998AJ....116.2965M,2009A&A...504..461F}. The energy from the accretion shock heats an area of the surrounding photosphere of the star to temperatures of the order of $20\;000$~K. In turn, this emits a hot black-body continuum, which veils the photospheric emission lines \citep{calvetgullbring,2000ApJ...544..927G}. The strength of the veiling in the UV and the optical wavelength range is one measure of the accretion rate. Some CTTS with fast rotation have been Doppler-imaged and hot spots on the surface can be seen \citep{1998MNRAS.295..781U,2005A&A...440.1105S}. Zeeman-Doppler imaging reveals that the Ca infrared triplet originates in a region of strong magnetic fields, which, in the magnetically funneled accretion model, are the footpoints of the accretion funnels \citep{2007MNRAS.380.1297D,2008MNRAS.386.1234D}.

CTTS and WTTS are also copious emitters of X-rays and UV radiation; \citet{1999ARA&A..37..363F} review the knowledge before X-ray grating spectroscopy with \emph{Chandra} and \emph{XMM-Newton}. In this article I will discuss coronal activity, which is common to both CTTS and WTTS (section~\ref{sect:corona}) and then summarize some observational characteristics in X-ray and UV spectroscopy, which set CTTS apart from main-sequence (MS) stars (section~\ref{sect:obs}) to discuss accretion (section~\ref{sect:accretion}), winds (section~\ref{sect:winds}) and stellar jets (section~\ref{sect:jets}) in the following. Section~\ref{sect:haebe} compares the properties of CTTS with HAeBes. I end with a short summary in section~\ref{sect:summary}.

\section{Classical T Tauri Stars}
\label{sect:ctts}
X-ray emission from T~Tauri stars (TTS) was discovered with the \emph{Einstein} satellite \citep{1981ApJ...243L..89F}. Starting with \emph{ROSAT}, X-ray surveys were used as a tool to identify young stars in star forming regions, e.g. in the Taurus molecular cloud \citep{1995A&A...297..391N}, the Chameleon star forming region \citep{1995A&AS..114..109A} and the Orion star forming cluster \citep{1996A&AS..119....7A}, because young stars have a high level of X-ray activity. This continues today, mostly with \emph{Chandra} because of its high imaging quality. One especially successful example is the Chandra Orion Ultradeep Project (COUP) with an exposure time of nearly 800~ks \citep{2005ApJS..160..319G}. However, the spectral information from these surveys remains poor by today's standards. Only the closest and brightest CTTS can be observed with high-resolution X-ray grating spectroscopy in a reasonable exposure time.

\subsection{Coronal activity}
\label{sect:corona}
For a long time the X-ray emission of CTTS was thought of as a scaled-up version of solar activity, because surveys of star forming regions show variability with a fast rise phase and a longer exponential decay in all types of TTS. CTTS and WTTS essentially share the same variability characteristics such as flare duration or the distribution of the flare luminosity, indicating that the same mechanism is responsible for the X-ray emission \citep[e.g.][]{2007A&A...468..463S,2008ApJ...688..418G,2008ApJ...688..437G}. The majority of flares is small, and as their energy increases, occurrence becomes rarer. In general, the number of flares $\textnormal{d}N$ in the energy interval $\textnormal{d}E$ follows a power-law with $\textnormal{d}N/\textnormal{d}E \propto E^{-\alpha}$. The value of the exponent $\alpha$ in, e.g., the Taurus molecular cloud is $2.5\pm0.5$ \citep{2007A&A...468..463S}, compatible with the solar value within the errors. This supports the idea that the majority of the X-ray emission on CTTS is coronal, just as the emission in WTTS or stars on the MS \citep[see][ for a review of stellar X-ray emission]{2009A&ARv..17..309G}.

However, WTTS are on average twice as bright as CTTS \citep{2001A&A...377..538S,2005ApJS..160..401P,2007A&A...468..425T}, so there must be some difference in the X-ray generation.

\subsection{Observational peculiarities}
\label{sect:obs}
The following two subsections describe spectral properties, discovered in high-resolution X-ray spectra, which set CTTS apart from WTTS or MS stars.

\subsubsection{He-like triplets}
X-ray grating spectra show the He-like ions of C, O, Ne, Mg and Si. These ions emit a triplet of lines, which consists of a resonance ($r$), an intercombination ($i$), and a forbidden line ($f$) \citep{1969MNRAS.145..241G,2001A&A...376.1113P}. The flux ratios of those lines are temperature and density-sensitive. The $R$- and $G$-ratios ($R = f/i$ and $G = (f+i)/r$) are commonly used to describe the triplet; for high electron densities $n_{\mathrm{e}}$ or strong UV photon fields the $R$-ratio falls below its low-density limit, because electrons are collisionally or radiatively excited from the upper level of the $f$ to the $i$ line, but the UV field of late-type CTTS is too weak to influence the $R$-ratio. The $G$-ratio is a temperature diagnostic of the emitting plasma. 

\begin{figure}
\resizebox{\hsize}{!}{\includegraphics{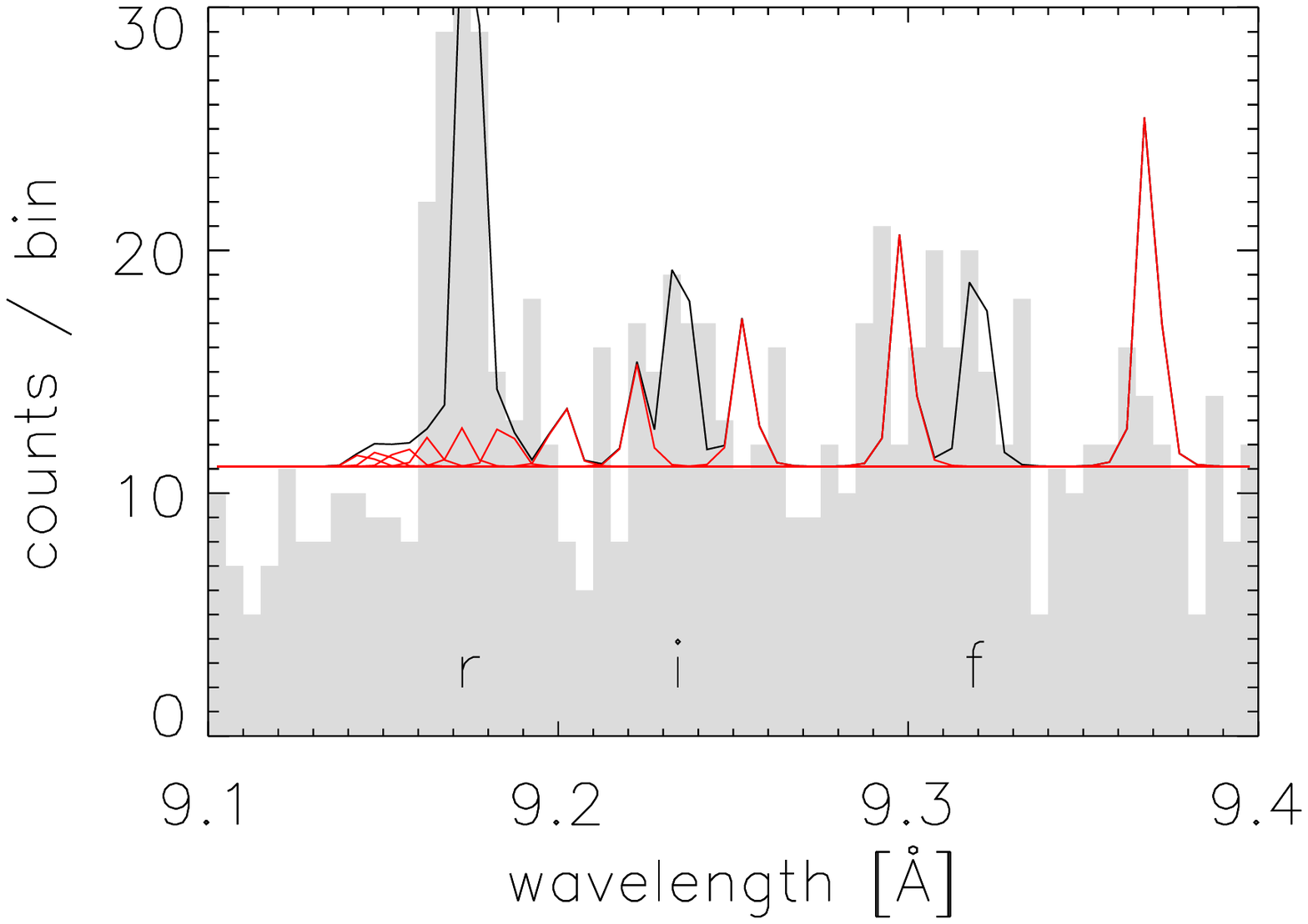}}
\resizebox{\hsize}{!}{\includegraphics{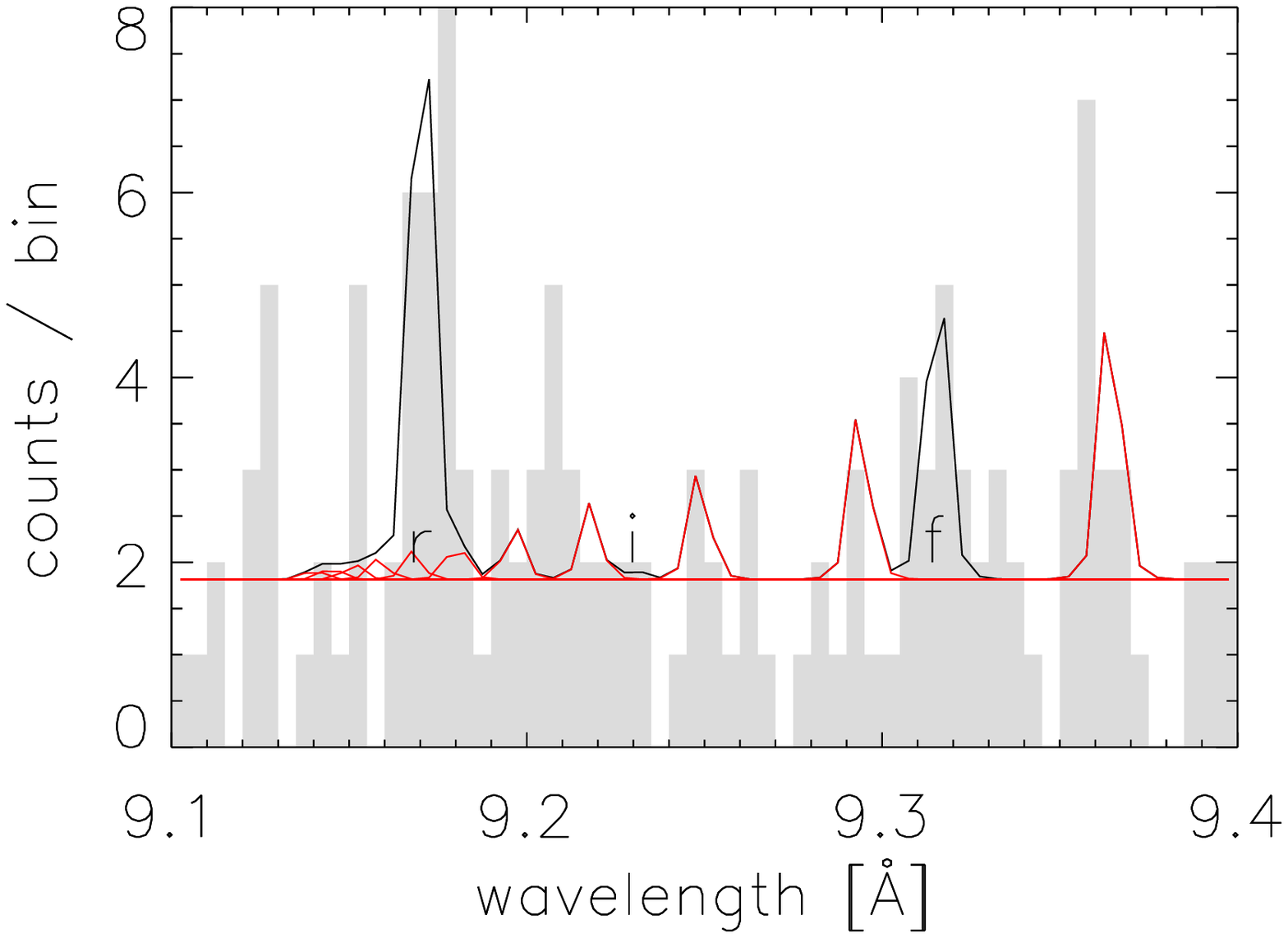}}
\caption{\ion{Mg}{xi} triplet of TW~Hya (top) and V4046~Sgr (bottom). The gray histogram shows the observed counts, the lines show the best-fit model. Extrapolated lines of the Lyman series a plotted in red/gray (color in electronic version only).}
\label{fig:nely}
\end{figure}

TW~Hya was the first CTTS to be observed by \emph{Chandra}/HETGS (for 50~ks). It showed an exceptional line ratio in the He-like triplets of \ion{O}{vii} and \ion{Ne}{ix}, which \citet{2002ApJ...567..434K} interpret as a signature of high density. TW~Hya has since been observed for 30~ks with \emph{XMM-Newton} \citep{twhya}, 120~ks with \emph{Chandra}/LETGS \citep{2009A&A...505..755R} and for 500~ks again with \emph{Chandra}/HETGS \citep{2010ApJ...710.1835B}. Together, this makes TW~Hya one of the targets with the longest exposure times in the history of X-ray spectroscopy. Other CTTS were observed with grating spectroscopy, too:  BP~Tau \citep{bptau}, V4046~Sgr \citep{v4046}, RU~Lup \citep{RULup}, MP~Mus \citep{2007A&A...465L...5A} and Hen~3-600 \citep{2007ApJ...671..592H} and they all show the same indication for high-densities that were first seen in TW~Hya. 
Figures~ \ref{fig:nely} and \ref{fig:accr} (middle and right column) show examples for the He-like triplets in CTTS. TW~Hya has the most extreme $f/i$ ratio observed so far, but also more typical CTTS like V4046~Sgr differ markedly from the line ratio found in typical WTTS such as TWA~4 \citep{2004ApJ...605L..49K} and TWA~5 \citep{twa5} and MS stars \citep{2004A&A...427..667N}, which are compatible with the low-density limit of the $f/i$ line ratio according to the CHIANTI 5.1 database \citep{CHIANTII,CHIANTIVII}.

There are exceptions to the high-densities in CTTS -- in the more massive, eponymous T~Tau itself the \ion{O}{vii} triplet is consistent with the coronal limit \citep{ttau}.

The \ion{O}{vii} triplet is free of blends and its $R$-ratio is sensitive to densities of $10^{11}-10^{12}$~cm$^{-3}$, but \emph{Chandra}/ACIS data has a very low effective area at this wavelength. The \ion{Ne}{ix} triplet is often blended with iron lines, particularly \ion{Fe}{xix} and \ion{Fe}{xx}. Strong sources provide sufficient signal in \emph{Chandra}/HEG, where most of these blends are resolved, but for lower fluxes or observations with \emph{XMM-Newton} these blends are difficult to remove, if the temperature structure is not very well known. Fortunately, most CTTS show an enhanced Ne abundance and a reduced Fe abundance, alleviating this problem to some extent (section~\ref{sect:accretion}).

The third interesting triplet is the \ion{Mg}{xi} He-like triplet with lines at 9.17~\AA{}, 9.23~\AA{} and 9.31~\AA{}. It can be blended by the higher members of the \ion{Ne}{x} Lyman series. This problem is somewhat more serious in CTTS than in other objects, because of the enhanced Ne abundance often found in CTTS. For a few active MS stars \citet{2004ApJ...617..508T} extrapolated the strength of the higher order \ion{Ne}{x} lines from the lower ones, which can be easily measured. They then fit the \ion{Mg}{xi} triplet, taking the line blends into account. \citet{2010ApJ...710.1835B} analyzed their \emph{Chandra}/HETGS observations and found that the blending is not important in this case.

For densities $<10^{15}$~cm$^{-3}$ the collision time is much longer than the radiative decay time in the Lyman series, thus all excited states decay radiatively and the relative strength of the unabsorbed \ion{Ne}{x} lines depends only on the temperature and the collision strength. I calculated the Lyman series up to Ly$\epsilon$ for a grid of temperatures with the CHIANTI 5.1 code \citep{CHIANTII,CHIANTIVII} and extrapolated that series with a geometric function. In the observed spectra I fit as many members of the Lyman series as possible with the CORA tool \citep{2002AN....323..129N}, which employs a maximum likelihood method, using a modified Lorentzian line profile with $\beta=2.5$ and keeping the line width fixed at the instrumental width of 0.02~\AA{} for the \emph{Chandra}/MEG. The fits to the \ion{Mg}{xi} triplet again use CORA. Here, the maximum likelihood is calculated for the sum of the extrapolated \ion{Ne}{x} lines, \ion{Mg}{xi} triplet and a constant, which represents a combination of background, continuum emission and unresolved lines. The likelihood is minimized with Powell's method by adjusting the individual flux of the triplet members, the constant and a wavelength offset. Both CTTS with archival \emph{Chandra}/HETGS observations, TW~Hya and V4046~Sgr, have very low fitted absorbing column densities for the accretion shock in the global fit (see section~\ref{sect:accr:obs} and \ref{sect:accr:fits}), thus the differential transmission between 9.1~\AA{} and 12.13~\AA{} is negligible.
Table~\ref{tab:nely} gives the line fluxes found and figure~\ref{fig:nely} shows the fit.
\begin{table*}
\caption{\ion{Ne}{ix} Ly series and \ion{Mg}{xi} triplet fluxes (errors are $1\sigma$ confidence intervals) \label{tab:nely}}
\begin{tabular}{lrrrrr}\hline
line & $\lambda$ & \multicolumn{2}{c}{TW Hya -- Chandra} & \multicolumn{2}{c}{V4046 Sgr -- Chandra}\\ 
 & \AA{} & cts & cts~s$^{-1}$~cm$^{-2}$ & cts & cts~s$^{-1}$~cm$^{-2}$ \\
\hline
\ion{Ne}{x} Ly$\alpha$ & 12.13 & $1345\pm40$ & $(76.\pm2.0)\times10^{-6}$ & $274\pm17$ & $(51.\pm3.0)\times10^{-6}$\\
\ion{Ne}{x} Ly$\beta$  & 10.24 & $ 247\pm17$ & $(8.3\pm0.6)\times10^{-6}$ & $ 57\pm 8$ & $(5.6\pm0.6)\times10^{-6}$\\
\ion{Ne}{x} Ly$\gamma$ &  9.71 & $  73\pm10$ & $(2.4\pm0.3)\times10^{-6}$ & $ 25\pm 6$ & $(2.4\pm0.6)\times10^{-6}$\\
\ion{Ne}{x} Ly$\delta$ &  9.48 & $  62\pm11$ & $(1.5\pm0.3)\times10^{-6}$ & $ 25\pm 8$ & $(2.0\pm0.7)\times10^{-6}$\\
\hline
\ion{Mg}{xi} r &  9.17 & $43.\pm8.6$ & $(8.7\pm1.7)\times10^{-7}$ & $10.\pm3.8$ & $(7.2\pm2.7)\times10^{-7}$\\
\ion{Mg}{xi} i &  9.23 & $16.\pm6.4$ & $(3.3\pm1.3)\times10^{-7}$ & $0.1\pm1.7$ & $(0.07\pm1.2)\times10^{-7}$\\
\ion{Mg}{xi} f &  9.31 & $16.\pm6.3$ & $(3.2\pm1.3)\times10^{-7}$ & $5.5\pm2.9$ & $(3.9\pm2.1)\times10^{-7}$\\
\hline
\end{tabular}
\end{table*}
The constant is the strongest contributor to the total flux. Some of the Ne lines are clearly overpredicted, e.g. at 9.36~\AA{} for TW~Hya, others match the observations within the (large) Poisson errors. In any case, the \ion{Mg}{xi} lines are not significantly blended by the Ne Lyman series. For TW~Hya, this confirms the finding of \citet{2010ApJ...710.1835B}.

The low-density limit of the $f/i$ ratio for \ion{Mg}{xi} is 2.5 for densities $<10^{12}$~cm$^{-3}$; the ratio drops below 0.2 for densities $>10^{14}$~cm$^{-3}$. Given the large statistical error on the line fluxes, neither the high-density nor the low-density limit can be excluded for TW~Hya or V4046~Sgr, but at least for V4046~Sgr a low-density scenario is more likely as the $i$ line is absent.

\subsubsection{Ionization balance}
To constrain the temperature and the origin of the plasma at low temperatures, which produces the anomalous $R$-ratios in the He-like triplets a line based method is preferable. Ideally, one would use the $G$-ratio of He-like triplets. Unfortunately, the temperature dependence of the $G$-ratio is weak and thus requires a high signal-to-noise ratio, which is only available for TW~Hya. \citet{2010ApJ...710.1835B} use the higher order resonance lines of \ion{O}{vii} for this purpose.

\begin{figure}
\resizebox{\hsize}{!}{\includegraphics{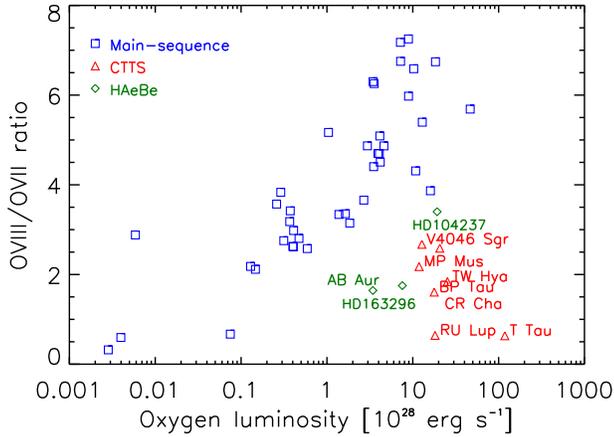}}
\caption{Flux ratio of \ion{O}{viii}~Ly$\alpha$ and \ion{O}{vii} He-like triplet, plotted over the sum of those emission lines. For comparison purposes MS stars from the \citet{2004A&A...427..667N} sample are shown. Color in electronic version only.}
\label{fig:o82o7}
\end{figure}

Alternatively, the balance of different ionization stages can be used as a temperature diagnostic.
\citet{RULup} and \citet{manuelnh} constructed a diagram, which shows the ratio of the \ion{O}{viii} to the \ion{O}{vii} lines and \citet{HD163296} added more stars (figure~\ref{fig:o82o7}). The flux ratios in this diagram are based on unabsorbed luminosities, where the flux has been corrected for the extinction, which was found in a global fit, using the absorption cross-sections from \citet{1992ApJ...400..699B}. This step is necessary, because the extinction varies with wavelength, so different correction factors apply for \ion{O}{viii} at 18.97~\AA{} and the \ion{O}{vii} triplet around 21.8~\AA{}.
All CTTS are found at the bottom right of the diagram at comparatively high total luminosities. This is a selection bias, because X-ray grating spectroscopy can only be performed for the brightest CTTS in reasonable exposure times. The distance to the TW~Hya association is about 57~pc, the next closest regions such as the Taurus-Auriga cloud or the $\rho$~Oph cloud are located at a distance of 130~pc. In contrast, the MS sample from \citet{2004A&A...427..667N} contains many closer stars and thus reaches down to lower luminosities. Still, the diagram shows an excess of \ion{O}{vii} emission in CTTS compared to MS stars of similar luminosity. The peak emission temperature for the \ion{O}{vii} lines is around 1-2~MK.

\begin{figure}
\resizebox{\hsize}{!}{\includegraphics{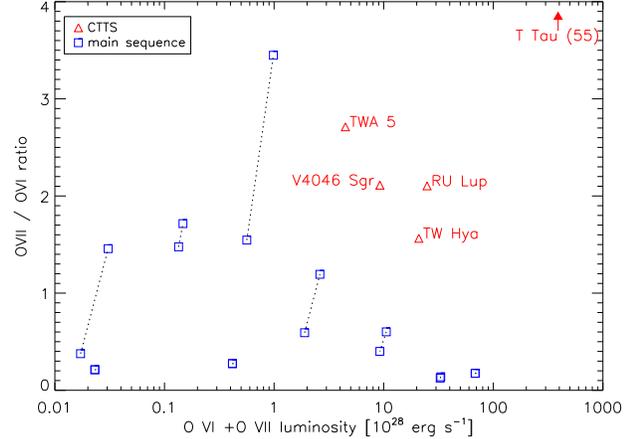}}
\caption{Flux ratio of \ion{O}{vii} He-like triplet to the \ion{O}{vi} doublet at 1032 and 1038~\AA{}, plotted over the sum of those emission lines. For comparison purposes MS stars are shown. If two X-ray observations are available in \citet{2004A&A...427..667N}, the figure shows two ratios calculated from those fluxes to give an estimate of variability. Color in electronic version only. }
\label{fig:o72o6}
\end{figure}

Following the same line of thought, the ratio of \ion{O}{vii} and \ion{O}{vi} emission will probe the cooler end of the plasma distribution. As no \ion{O}{vi} emission lines are found in the X-ray range, they have to be taken from non-simultaneous observations with the \emph{Far Ultraviolet Spectroscopic Explorer (FUSE)}, which covers the wavelength range of the \ion{O}{vi} doublet at 1032 and 1038~\AA{}. The \ion{O}{vi} fluxes for MS stars are taken from \citet{2002ApJ...581..626R} and \citet{2005ApJ...622..629D}, those for CTTS from \citet{FUSElineforms}. The observed \emph{FUSE} fluxes are dereddened with the reddening law of \citet{1989ApJ...345..245C}. The X-ray and UV observations are taken up to a few years apart, so the line ratios could be influenced by variability. To asses this figure~\ref{fig:o72o6} shows two flux ratios for those MS stars which have been observed multiple times in X-rays \citep{2004A&A...427..667N}. Variability dominates over the formal measurement uncertainty, which is around 10\% for most stars. 
Again, the CTTS are separated from MS stars in this diagram, although less significant than in figure~\ref{fig:o82o7}. This time, they lie above the MS stars, indicating that they have extra \ion{O}{vii} emission compared to MS stars with similar luminosity in the oxygen lines. So, CTTS are more luminous in \ion{O}{vii} compared to both hotter (\ion{O}{viii}) and cooler (\ion{O}{vi}) plasma than MS stars.  Given that CTTS and WTTS seems to have very similar coronal emission, CTTS even appear to be hotter, it is unlikely, that this is caused by CTTS being \emph{under}luminous in \ion{O}{viii} and \ion{O}{vi}. A much better explanation for this excess of soft X-ray flux is an extra emission component at 1-2~MK (the formation temperature of the \ion{O}{vii} He-like triplet) that is present in CTTS, but not in MS stars.

\subsection{Accretion}
\label{sect:accretion}
One promising candidate for this extra emission is the accretion shock on the star. Here, an overview of accretion shock models is presented in section~\ref{sect:accr:models}, then a sample of CTTS it described, where high resolution X-ray spectra indicate the presence of an accretion component (section~\ref{sect:accr:obs}). Section~\ref{sect:accr:fits} shows model fits to this dataset. A comparison between our accretion shock model \citep{acc_model} and similar models in the literature is given in section~\ref{sect:accr:comp}.

\subsubsection{Accretion models}
\label{sect:accr:models}
The inner hole in the accretion disk is at least a few stellar radii wide, and thus the accretion impacts on the star close to free-fall velocity $v_0$. For a star with mass $M_*$ and radius $R_*$ this is:
\begin{equation} v_0=\sqrt{\frac{2GM_*}{R_*}} =  600 \sqrt{\frac{M_*}{M_\odot}}\sqrt{\frac{R_\odot}{R_*}} \frac{\textnormal{km}}{\textnormal{s}}\ . \label{eqn:infallvel} \end{equation}
Typically, CTTS have masses comparable to the Sun and radii between $R_*=1.5\;R_{\sun}$ and  $R_*=4\;R_{\sun}$ \citep{2003ApJ...597L.149M}, because they have not yet finished their contraction. This gives infall velocities in the range 300-500~km~s$^{-1}$. Not all kinetic energy is converted into thermal energy, because the total momentum needs to be conserved. In strong shocks the post-shock velocity $v_1$ is given by 
$$v_1 = \frac{1}{4} v_0\ ,$$ 
where $v_0$ is the pre-shock velocity. Particle number conservation then demands the following relation for the pre-shock particle number density $n_0$ and the post-shock number density $n_1$: $$n_1 = 4 n_0\ .$$ The pre-shock pressure is completely dominated by the ram pressure of the infalling material, which drops dramatically across the shock front. In a quasi-equilibrium state, this is compensated by thermal pressure on the post-shock side and this leads to an expression for the post-shock temperature $T_1$:
$$ T_1=\frac{3}{16}\frac{\mu m_{\mathrm{H}}}{k} v_0^2,$$
where $\mu$ is the particle mass number, averaged over ions and electrons, and $m_{\mathrm{H}}$ is the mass of a hydrogen atom. $k$ is the Boltzmann constant. For infall velocities close to the free-fall velocity this equation predicts extra emission at 1-2~MK. Strictly speaking, mostly the ions are heated in the strong shock, and it takes several mean-free path lengths to transfer heat to the electron gas. However, non-equilibrium ionization and temperature have a negligible effect on the final spectrum \citep{acc_model}.
\begin{figure}
\resizebox{\hsize}{!}{\includegraphics{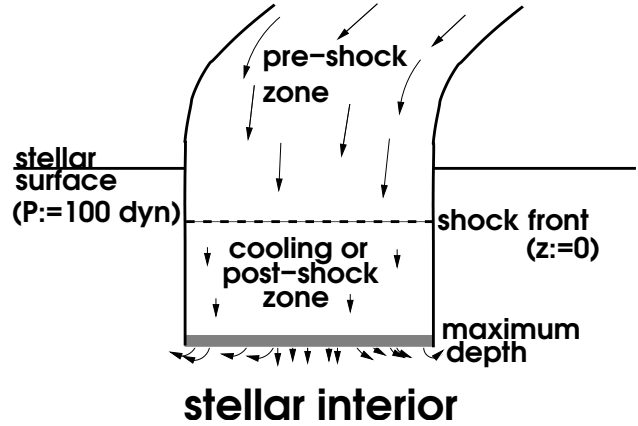}}
\caption{Sketch of the accretion shock geometry. This sketch is not to scale. Matter falls in from the accretion streams and close to the stellar surface an accretion shock develops. In the post-shock zone the plasma cools via radiation. The stellar surface is arbitrarily represented at pressure $p=100$~dyn.}
\label{fig:funnel}
\end{figure}
After plasma passes through the accretion shock, it cools down via radiation. If the magnetic field of the accretion funnels is strong enough, all motion perpendicular to the field lines is suppressed and the problem can be simulated in 1D (sketch in figure~\ref{fig:funnel}). \citet{calvetgullbring} presented a grid of such models. They find that the radiation from the accretion shock heats the photosphere below the shock to about $20\;000$~K. Their model fits the hydrogen continuum in the UV. \citet{lamzin} calculated the expected X-ray flux from the post-shock cooling zone. These simulations were updated by \citet{acc_model} to resolve individual lines, especially in the He-like triplets. CTTS definitely show signs of coronal activity (section~\ref{sect:corona}), so observed X-ray spectra have to be fitted with a combination of models which account for accretion and coronal contribution. 
\citet{2005CSSS.519.D} suggested that the shock might be buried in the photosphere, so that most of the X-ray radiation is absorbed by the surrounding photospheric gas. However, the fact that we observe high densities shows that at least some radiation escapes. While the photosphere would be a gray absorber, which affects all lines of the triplets simultaneously, emission from deeper layers of the post-shock cooling zone could experience line absorption. There are some hints to optical depth effects in the X-ray spectra \citep{2009A&A...507..939A}, but for most stars the $G$-ratios are close to the expected values. This excludes large optical depth effects, because the optical depth of the resonance line should be several orders of magnitude larger than in the forbidden or the intercombination line. 

I calculated a grid of shock model with the code of \citet{acc_model} for infall velocities between 300~km~s$^{-1}$ and 1000~km~s$^{-1}$ in steps of 100~km~s$^{-1}$ and infall densities between $10^{10}$~cm$^{-3}$ and $10^{14}$~cm$^{-3}$ in five logarithmic steps. The emission for all ions of C, N, O, Ne, Mg, Si, S and Fe is calculated separately from the continuum emission and the remaining elements, so that these abundances can be fit individually. However, large changes in the abundance would significantly alter the cooling function and thus the thermal structure of the shock. Simulations with different abundances found in the literature for CTTS were performed, and in this range no significant change was observed. The model grid is provided as a table model. This table model is available at \texttt{http:\/\/www.hs.uni-hamburg.de\/DE\/Ins\/Per\/Guenther\/shock\_model\/}.

This basic shock interpretation has been extended by \citet{2010ApJ...710.1835B}, who found a lower density and less absorption for \ion{O}{vii} than for \ion{Ne}{ix} in TW~Hya, although the \ion{O}{vii} triplet is formed at lower temperatures and therefore deeper in the post-shock cooling zone. They conclude, that most of the \ion{O}{vii} emission does not originate in the accretion shock itself, but in plasma that has been heated by the shocked material. This situation can no longer be described in 1D. Simulations of the accretion shock region in more dimensions have been performed by \citet{2010A&A...510A..71O}. They show the flow to be well constrained for strong magnetic fields, but weaker fields cannot hold the plasma and mass flows sideways, thus changing the temperature and density profiles in the shock.

There is no a-priory reason to expect the shocks to be stable over time. \citet{2008MNRAS.388..357K} and \citet{2008A&A...491L..17S} both looked at this issue and predicted sub-second oscillations of the shock front, albeit at very different time scales, mostly due to the vastly different densities they assume in the shock. Observationally, this has not been found, neither in X-rays \citep{2009ApJ...703.1224D} nor in the optical response \citep{TWHyalc}. This does not rule out fast oscillations of the shock front, it just requires the accretion along different field lines to oscillate independently with slightly different time constants.

\subsubsection{Observations and fitting}
\label{sect:accr:obs}
\begin{table*}
\caption{Observations\label{tab:obs}}
\begin{tabular}{llrll}\hline
star & satellite & exp. time & obs date & ObsID\\ 
 &  & [ks] & \\
\hline
TW Hya & XMM-Newton & 30  & 2001-07-09 & 0112880201\\
TW Hya & Chandra    & 490 & 2007-02-15 & 7435, 7436, 7437, 7438\\
BP Tau & XMM-Newton & 130 & 2004-08-15 & 0200370101\\
MP Mus & XMM-Newton & 110 & 2006-08-19 & 0406030101\\
V4046 Sgr & Chandra & 150 & 2006-08-06 & 5423, 6265\\
\hline
\end{tabular}
\end{table*}
Table~\ref{tab:obs} gives a list of X-ray observations of CTTS, where previous authors have found significant deviations from the low-density limit in the He-like triplets.

For observations split over several orbits, the day of the first exposure and the summed exposure time is given in the table. The \emph{XMM-Newton} data was retrieved from the archive and processed with the standard \emph{XMM-Newton} Science Analysis System (SAS) software, version 10.0 \citep{2004ASPC..314..759G} with all standard selection criteria to filter out background contamination.
\emph{Chandra} data was retrieved from the archive and processed with CIAO 4.3 \citep{2006SPIE.6270E..60F} to obtain CCD spectra; the grating spectra were taken directly from TGCat \footnote{http://tgcat.mit.edu}. Positive and negative first-order spectra were merged, then spectra from different orbits were combined, but HEG and MEG are kept separate. All CCD spectra were binned to 15 counts per bin, but grating data only to 5 counts per bin, because the density information in the triplets, which often contain only very few counts, would be destroyed by a coarser binning.

\subsubsection{A sample of CTTS}
\label{sect:accr:fits}
For small count numbers a $\chi^2$ statistic is no longer applicable, instead the fit was done using the C-statistic. Still, the CCD spectra with their higher count rates tend to dominate the statistic. For \emph{XMM-Newton} data only the MOS1 was fitted and for the \emph{Chandra} spectra only the ACIS zeroth order of one of the available exposures. In this way, the cool shock component, which is mostly constrained by the information from the gratings is effectively fitted, otherwise the optimization would prefer small improvements in the coronal components at the prize of systematic deviations in the low energy region because the hot components cause higher CCD count rates. The C-statistic takes the proper Poisson uncertainties in bins with small count number into account, but it does not provide a goodness-of-fit.

X-ray spectra can be used to fit relative abundances of metals, but it is very difficult to obtain absolute abundances. Thus, the abundance of oxygen is fixed at solar \citep{1998SSRv...85..161G} for all models in this section, i.e. all abundances are given relative to oxygen. The abundances of C, N, Ne, Mg, Si and Fe are fitted independently. The abundance of S is coupled to Fe, all other abundances are fixed at solar values. Because \emph{Chandra}/ACIS has low effective areas at longer wavelengths, where the lines of C and N are observed, I fixed the abundance of those two elements at solar values for all \emph{Chandra} data sets and, additionally, the absorbing column density for TW~Hya in the \emph{Chandra} data was fixed to the value found in the \emph{XMM-Newton} observation.

\begin{figure*}
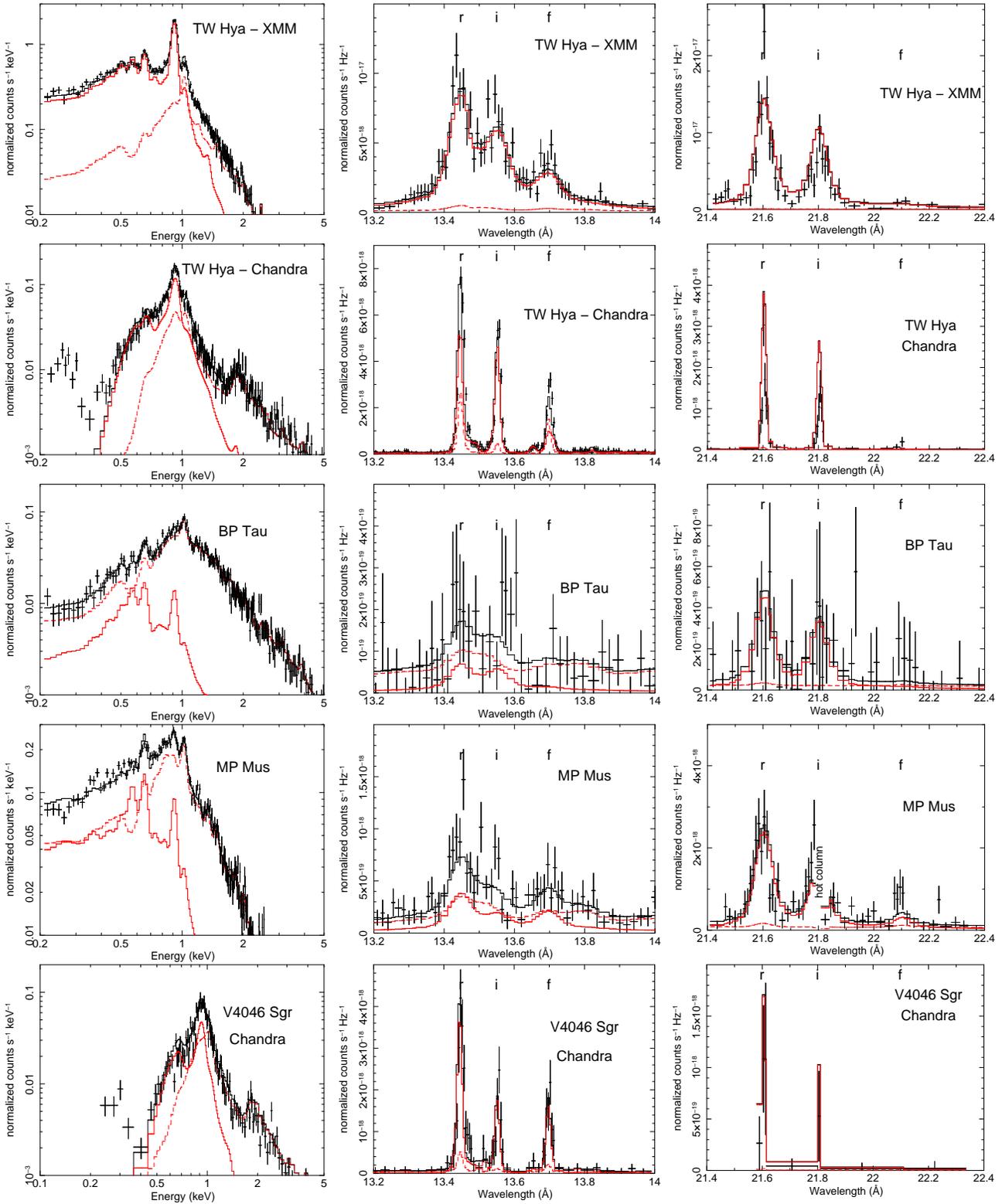

\resizebox{0.33\hsize}{!}{\includegraphics[angle=-90]{TWHyaXMMCCD}}
\resizebox{0.33\hsize}{!}{\includegraphics[angle=-90]{TWHyaXMMNe}}
\resizebox{0.33\hsize}{!}{\includegraphics[angle=-90]{TWHyaXMMO}}
\resizebox{0.33\hsize}{!}{\includegraphics[angle=-90]{TWHyaChandraCCD}}
\resizebox{0.33\hsize}{!}{\includegraphics[angle=-90]{TWHyaChandraNe}}
\resizebox{0.33\hsize}{!}{\includegraphics[angle=-90]{TWHyaChandraO}}
\resizebox{0.33\hsize}{!}{\includegraphics[angle=-90]{BPTauCCD}}
\resizebox{0.33\hsize}{!}{\includegraphics[angle=-90]{BPTauNe}}
\resizebox{0.33\hsize}{!}{\includegraphics[angle=-90]{BPTauO}}
\resizebox{0.33\hsize}{!}{\includegraphics[angle=-90]{MPMusCCD}}
\resizebox{0.33\hsize}{!}{\includegraphics[angle=-90]{MPMusNe}}
\resizebox{0.33\hsize}{!}{\includegraphics[angle=-90]{MPMusO}}
\resizebox{0.33\hsize}{!}{\includegraphics[angle=-90]{V4046CCD}}
\resizebox{0.33\hsize}{!}{\includegraphics[angle=-90]{V4046Ne}}
\resizebox{0.33\hsize}{!}{\includegraphics[angle=-90]{V4046O}}
\caption{Spectra of CTTS with fitted model overplotted. The panels show spectra in CCD resolution (left) and grating data for the \ion{Ne}{ix} (middle) and \ion{O}{vii} (right) He-like triplets. Best-fit models are overplotted, the contribution of corona (red/gray dashed) and shock (red/gray) is shown individually. In the He-like triplets most emission comes from the shock. The wavelength of the $r$, $i$ and $f$ line is labeled. The second and fifth row show \emph{Chandra} data, which has narrower lines in the grating spectra, but little effective area at lower wavelength in the CCD spectra.}
\label{fig:accr}
\end{figure*}

Fits to the CTTS sample were obtained with XSPEC~12.6 \citep{1996ASPC..101...17A}, fitting four model components: Three emission components (one shock model and two optically thin thermal APEC models, which represent the corona) and one cold photo-absorption component.
Figure~\ref{fig:accr} shows the fitted low-resolution spectra (left panels) and also the fits to the He-like ions \ion{Ne}{ix} (middle panels) and \ion{O}{vii} (right panels). The results are summarized in table~\ref{tab:accr_fits}.The panels show the contribution of shock and corona independently. The corona is responsible for the hot emission, which cannot be explained by the accretion shock, because such high temperature would require infall velocities above the free-fall velocity. The He-like triplets are predominantly formed at lower temperatures in the accretion shock. This can be immediately seen from the strong $i$ line. The corona is in the low density limit and thus its $f/i$ ratio is high in contrast to the observations. Therefore, the $i$ line has to be formed in the shock, which automatically requires the shock to contribute most of the emission in  the $r$ and $f$ lines as well.

Table~\ref{tab:accr_fits} only shows the statistical uncertainties for the fit, not any systematic model uncertainties. The plasma temperatures e.g. just represent the average plasma properties. The small statistical uncertainties on the temperature do not mean that the plasma temperature distribution is bimodal with very narrow peaks. Fits which prescribe e.g. polynomial emission measure distributions are also possible.

\begin{table*}
\caption{Fit results (errors are statistical only and give 90\% confidence intervals) \label{tab:accr_fits}}
\begin{tabular}{lrrrrr}\hline
 & TW Hya & TW Hya & BP Tau & MP Mus & V4046 Sgr\\
 & \emph{XMM-Newton} & \emph{Chandra} & \emph{XMM-Newton} & \emph{XMM-Newton} & \emph{Chandra}\\ 
\hline
\multicolumn{6}{c}{absorbing column density}\\
\hline
$N_{\mathrm{H}}$ [$10^{20}$~cm$^{-2}$] & $5.2^{+0.6}_{-0.4}$ & =5.2 & $14\pm4$ & $2\pm1$ & $6\pm3$\\
\hline
\multicolumn{6}{c}{abundances relative to O}\\
\hline
C  & $1.5\pm0.2$        & =1            &$2.2_{-1.7}^{+2.4}$& $1.4\pm0.4$ & =1\\
N  & $3.0\pm0.3$        & =1            &$2.2 \pm1.2$       & $1.6\pm0.4 $& =1 \\
Ne & $7.6^{+3.2}_{-2.6}$ & $3.2\pm0.2$   &$1.7\pm0.5$        & $1.8\pm0.2$ & $3.5\pm0.4$\\
Mg & $1.0\pm0.3$        & $0.36\pm0.05$ &$0.3\pm0.3$        & $0.8\pm0.2$ & $0.5\pm0.1$\\
Si & $0.8\pm0.3$        & $0.5\pm0.1$   &$0.2\pm0.2$        & $0.6\pm0.2$ & $0.6\pm0.1$\\
Fe & $0.9\pm0.1$        & $0.42\pm0.03$ &$0.2\pm0.1$        & $0.5\pm0.1$ & $ 0.46\pm 0.06$\\
\hline
\multicolumn{6}{c}{coronal components}\\
\hline
k$T_1$ [keV]                 & $0.69_{-0.02}^{+0.04}$ & $0.36\pm0.01$ &$0.7\pm0.1$ & $0.63\pm0.03$ & $ 0.73\pm 0.04$\\
$VEM_1$ [$10^{51}$cm$^{-3}$] & $7.2\pm1.2$            & $27\pm2$    &$43\pm2$ & $83\pm15$ & $28\pm5$\\
k$T_2$ [keV]                 & $1.7_{-0.1}^{+0.2}$     & $2.0\pm0.05 $&$2.2\pm0.2$ & $2.2\pm0.1$  & $2.2\pm0.2$\\
$VEM_2$ [$10^{51}$cm$^{-3}$] & $19\pm1.5$             & $37\pm1    $&$100\pm10$ & $111\pm5$     & $38\pm3$\\
\hline
\multicolumn{6}{c}{shock properties}\\
\hline
$v_0$ [km~s$^{-1}$]             & $500\pm5$           & $504\pm3$   &$440_{-25}^{+70}$ & $510\pm10$ & $500_{-50}^{+10}$\\
$\log(n_0)$ [cm$^{-3}$]         & $11.9^a$        & $13.0^a$&$12.9^a$& $11.1\pm0.2$ & $11.2^a$\\
$A$ [cm$^2$]                    & $6.5\times10^{19}$  & $4\times10^{18}$ & $4\times 10^{20}$&$4\times10^{20}$ & $3\times10^{20}$\\
$\dot M$ [$M_{\sun}$~yr$^{-1}$] & $7 \times 10^{-11}$ & $6\times10^{-11}$& $4\times10^{-9}$& $6\times10^{-11}$ &$6\times10^{-11}$\\
\hline
\multicolumn{6}{c}{X-ray mass accretion rates from literature}\\
\hline
$\dot M$ [$M_{\sun}$~yr$^{-1}$] & $2\times10^{-11}$ (1) & & $9\times10^{-10}$ (2) & $2\times10^{-11}$ (3)  & $3\times10^{-11}$ (4)\\
$\dot M$ [$M_{\sun}$~yr$^{-1}$] & $2\times10^{-10}$ (5) & & &$7.7\times10^{-11}$ (6)\\
\hline
\end{tabular}\\
(a) The model grid uses density steps of 1 in log-space. Here, the interpolation error between two models is much larger than the statistical uncertainty.
(1) \citet{twhya}; (2) \citet{bptau}; (3) \protect{\citet{2007A&A...465L...5A}}; (4) \citet{Bonn06}; (5) \citet{acc_model}; (6) \protect{\citet{2008A&A...491L..17S}}
\end{table*}
The mass accretion rate $\dot M$ is related to the accretion spot size $A$, the infall velocity $v_0$ and the density $n_0$ as:
\begin{equation}
\dot M = \mu m_{\mathrm{H}} n_0 v_0 A \ .
\end{equation}
For most stars the signal in the He-like ions is relatively weak and the statistical error on the density is large. The error in $n_0$ and $A$ is then correlated such that a small accretion spot with a high density or a large accretion spot with a low density are both possible, if they have the same mass accretion rate and thus the same total luminosity. So, while $n_0$ and $A$ are uncertain, $\dot M$ is still a well-determined quantity. Fitted mass accretion rates are given in table~\ref{tab:accr_fits}.

The sample was selected to contain only stars with high densities. The prime example is TW~Hya, where \citet{2002ApJ...567..434K} observed $10^{13}$~cm$^{-3}$, \citet{twhya} the same, \citet{2009A&A...505..755R} $10^{12}$~cm$^{-3}$ and \citet{2010ApJ...710.1835B} between $6\times10^{11}$~cm$^{-3}$ and $3\times10^{12}$~cm$^{-3}$ for \ion{O}{vii} and \ion{Ne}{ix} respectively.
In BP~Tau \citet{bptau} find $3\times10^{11}$~cm$^{-3}$, \citet{v4046} give $3\times10^{11}-10^{12}$~cm$^{-3}$ for V4046~Sgr and \citet{2007A&A...465L...5A} found $5\times10^{11}$~cm$^{-3}$ in MP~Mus. This is largely consistent with the post-shock values expected from the pre-shock densities given in table~\ref{tab:accr_fits}.

If, however, the cool emission is formed only partly in a high-density accretion shock and partly in a low-density corona, the $f/i$ ratio in the He-like triplet can indicate medium densities. In this case, the $\dot M$ is overestimated. 
One further problem of the model chosen in this paragraph shows up in TW~Hya. \citet{2010ApJ...710.1835B} found that the accretion shock is less absorbed than the corona in TW~Hya and the absorbing column density differs even between \ion{Ne}{ix} and \ion{O}{vii}. Also, \ion{Ne}{ix} indicates a larger density than \ion{O}{vii}, although is should be formed higher in the accretion shock. All this is not reproduced by the simple model used here and figure~\ref{fig:accr} shows that \ion{Ne}{ix} is fitted very well for TW~Hya, while \ion{O}{vii} is overpredicted.

Unfortunately, TW~Hya is the only CTTS with such a high signal-to-noise in the spectrum that these differences to the model can be seen. Similar shortcomings as in TW~Hya might exist for the other CTTS but would be hidden in the noise. The best I can do to compare the mass accretion rates for several CTTS consistently is to fit all stars in the sample with a consistent model.

The table also shows, that the Ne abundances are enhanced in all CTTS, while the abundances of Fe, Si and Mg are reduced. \citet{twhya} put forward the idea that Fe, Si and Mg condense on grains and settle, but the noble gases cannot and are accreted onto the star. However, Ne also has a higher first ionization potential and in all active stars the coronal abundance of those elements is enhanced (IFIP effect) \citep[see the review by][]{2009A&ARv..17..309G}. Even for TW~Hya the signal is not strong enough to determine the Ne abundance in the shock and the corona independently, so any one or both of the above scenarios might be important.

In most cases, the corona is described by one temperature component with $kT\approx0.7$~keV and another with $kT\approx2$~keV; the volume emission measures $VEM$ of the corona vary over the sample. That is not surprising as several of the lightcurves contain strong flares which cause higher temperatures and higher emission measures for the duration of the flare.

The fitted values for the infall velocity in all stars are compatible with estimates for the free-fall velocity. The normalization of the shock models is proportional to the total mass flux, with is around $6\times 10^{-11}M_{\sun}$~yr$^{-1}$ for TW~Hya, MP~Mus and V4046~Sgr. TW~Hya and MP~Mus are both relatively old CTTS, so a smaller mass accretion rate here is expected. The mass accretion rate of BP~Tau is two orders of magnitude stronger.

\subsubsection{Comparison of mass accretion rates}
\label{sect:accr:comp}
The shock models fitted in the previous section agree with other estimates in the literature, that are also based on X-ray spectra, within a factor of 2-4 (table~\ref{tab:accr_fits}). Some of those estimates are far simpler \citep{twhya,bptau,2007A&A...465L...5A}. They just use the fitted volume emission measure and density with a single value for the cooling function and some rough estimate for the depth of the post-shock cooling zone. Still, the results are comparable. In the case of TW~Hya the mass accretion rates of both observations agree. Again, this shows that the fitting of the mass accretion rates is relatively robust, but the density should be determined from line fitting and not with a global fit on binned spectra. 

However, all X-ray determined mass accretion rates are one or two orders of magnitude smaller than measurements obtained with the UV flux or the optical veiling \citep{2010arXiv1011.5915C}. It is unknown what causes this effect. One possibility are inhomogeneous accretion spots. Either only a small part of the accretion stream impacts at free-fall velocity and the remaining mass accretes at a lower velocities or parts of the accretion spots are hidden by complete absorption. Although some hydrodynamical simulations of accretion streams show inhomogeneous impact velocities \citep{2004ApJ...610..920R,2007MNRAS.374..436L}, it remains unclear which physical mechanism reduces the speed and where the corresponding gravitational energy is lost. Also, \citet{2009A&A...507..939A} find no resonant scattering in the \ion{O}{vii} lines in TW~Hya, thus is seems unlikely that absorption can explain the small mass accretion rate found.

\subsection{Winds}
\label{sect:winds}
Many, if not all, CTTS have outflows \citep[a review is given by][]{2007prpl.conf..215B}, but the physical driving mechanism is unknown.  Theoretical models propose a variety of stellar winds \citep{1988ApJ...332L..41K,2005ApJ...632L.135M}, X-winds \citep{1994ApJ...429..781S} and disc winds \citep{1982MNRAS.199..883B,2005ApJ...630..945A}. Winds, which could be of stellar origin or come from the disk, are observed over a wide wavelength range from radio to the UV \citep[e.g.][]{2000AJ....119.1881A,2001ApJ...551.1037B,2004AstL...30..413L,2005ApJ...625L.131D,2006ApJ...646..319E}. Some CTTS also have highly collimated jets, which are discussed in more detail in section~\ref{sect:jets}. The dynamics of the gas around CTTS are measured by UV spectroscopy with \emph{HST/GHRS}, \emph{HST/STIS} and \emph{FUSE} \citep{2002ApJ...566.1100A,2002ApJ...567.1013A, 2002ApJ...572..310H, 2005AJ....129.2777H, 2006ApJS..165..256H}. The best-studied CTTS is TW Hya, where \citet{2005ApJ...625L.131D} fit the asymmetric line profile of the \ion{O}{vi}~1032~\AA{} line with a Gaussian with the centroid matching the stellar rest-frame. They explain the missing flux on the blue side of the line by a spherically-symmetric and smoothly-accelerated hot wind. However, \citet{2007ApJ...655..345J} argue that this model is incompatible with \emph{HST/STIS} observations, especially for the \ion{C}{iv} 1550~\AA{} doublet.
\begin{figure}
\resizebox{\hsize}{!}{\includegraphics{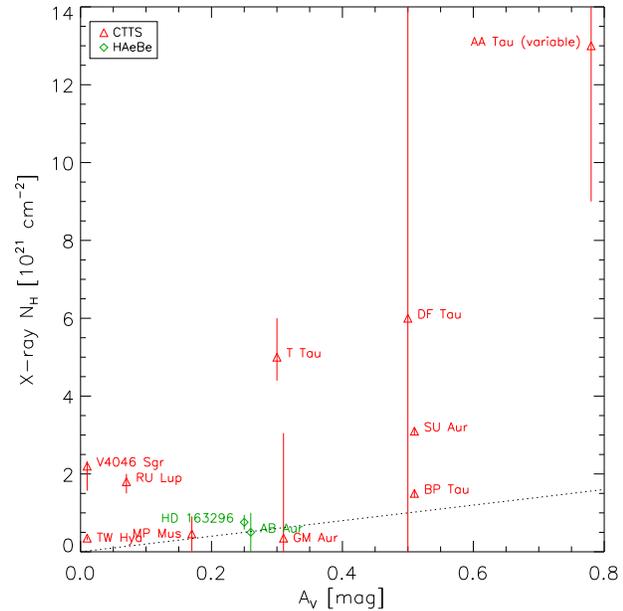}}
\caption{Optical reddening and X-ray absorbing column density (with 90\% confidence error bars). The dotted line shows the interstellar gas-to-dust ratio. DM~Tau and GM~Aur have \emph{ROSAT}/PSPC spectra only, so the error bars are much larger.}
\label{fig:nhav}
\end{figure}
The existence of a hot wind in TW Hya, therefore, remains an open issue. \citet{FUSElineforms} extracted line profiles for all CTTS in the \emph{FUSE} archive and find line centroids to be shifted between -170~km~s$^{-1}$ and +100~km~s$^{-1}$. The blue-shifted emission is likely caused by shocks in outflows from the CTTS.

Figure~\ref{fig:nhav} compares the gas column density as measured by X-rays with the optical reddening, which is mainly caused by dust \citep[see][ and references therein for data sources]{FUSElineforms}. The optical reddening is notoriously difficult to measure in accreting systems, but at least in some cases, notably V4046~Sgr, RU~Lup and T~Tau, the gas absorption is much higher than expected from an interstellar gas-to-dust ratio \citep{2003A&A...408..581V} and the optical reddening. These are also the stars with blue-shifted UV emission. It is possible that the same outflows, which provide the UV emission lines also act as dust-depleted absorbers for the stellar light.

\subsection{Jets}
\label{sect:jets}
In addition to wide-angle winds CTTS can also drive highly collimated jets \citep{1995RMxAC...1....1R,2004ApJ...604..758C,2005ApJ...626L..53G}, but DG~Tau is the only CTTS where X-ray emission has been found from the jets. Other X-ray jets like HH~2 \citep{2001Natur.413..708P} or HH~154 \citep{2002A&A...386..204F,2003ApJ...584..843B,2006A&A...450L..17F} have younger and more embedded driving sources.
\begin{figure}
\resizebox{\hsize}{!}{\includegraphics{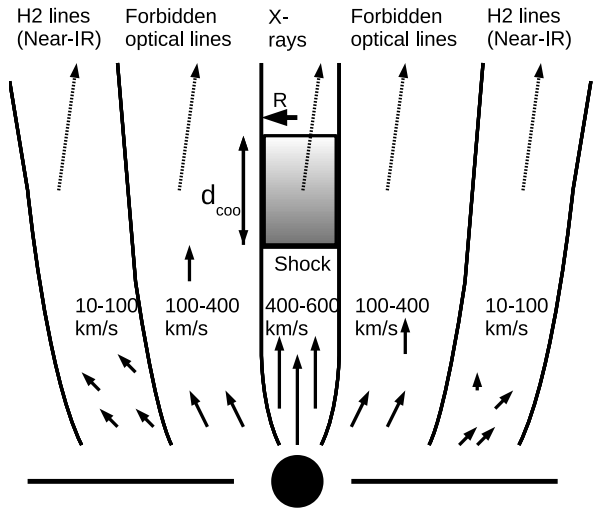}}
\caption{Sketch of a cut through the outflow from a CTTS (not to scale). The inner components of the outflow are more collimated and faster and thus heat to higher temperatures when shocked.}
\label{fig:jetsketch}
\end{figure}
Usually, the inner components of these jets are faster and denser (figure~\ref{fig:jetsketch}). \citet{2000ApJ...537L..49B} used the \emph{HST} to resolve the jet of DG Tau in several long-slit observations. The emission of gas faster than 200~km~s$^{-1}$ is mostly confined to the inner slit, corresponding to a radius of 15~AU. The ratios of [\ion{O}{i}], [\ion{N}{ii}] and [\ion{S}{ii}] give a lower limit on the gas density of $10^4$~cm$^{-3}$. This component seems to be surrounded by slower moving gas with lower densities of the order $10^3-10^4$~cm$^{-3}$. On larger scales an even cooler outflow, with an opening angle of 90\degr{} is seen in molecular hydrogen \citep{2004A&A...416..213T}. \citet{1997A&A...327..671L} estimate a mass loss rate of $6.5\cdot 10^{-6}$~M$_{\sun}$~yr$^{-1}$ assuming shock heating in the gas. \citet{1995ApJ...452..736H} obtain $3\cdot 10^{-7}$~M$_{\sun}$~yr$^{-1}$. The kinematics of the gas can be calculated from line shifts and proper motion, accounting for the inclination of the jet to the plane of the sky. Optical and IR lines are blue-shifted up to deprojected velocities of 600~km~s$^{-1}$ \citep{2000A&A...356L..41L,2000ApJ...537L..49B,2003ApJ...590..340P} and the proper motion of the knots in the jets is 0\farcs28~yr$^{-1}$, which translates into a deprojected velocity of 300~km~s$^{-1}$ \citep{2003ApJ...590..340P}.

In this context, the X-ray emission from the jet can be understood as shock-heated plasma from the densest and innermost component of the outflow \citep{dgtau}. Only $10^{-3}$ of the total mass outflow is required to shock at the velocities observed in the fastest jet component to explain the observed X-ray spectra and emission measures. Given the density from the forbidden emission lines and the temperature for the X-ray spectrum the cooling length $d_{\mathrm{cool}}$ can be calculated with a model very similar to the accretion shock model. The total intensity determines the volume emission measure of the plasma and with the density and $d_{\mathrm{cool}}$ this yields the area of the shock. For DG~Tau the estimated dimensions are only a few AU, if all X-ray emission is produced in a single shock (figure~\ref{fig:jetsketch}). Because of the small dimensions, a shock in the innermost component does not necessarily disturb the flow in the outer layers, which are resolved with the \emph{HST}. So far, this model is compatible with all available observations.

\section{Herbig Ae/Be stars}
\label{sect:haebe}
Herbig~Ae/Be stars (HAeBes) are in a similar evolutionary state as CTTS, but they have spectral type A or B. Due to their higher mass their evolution progresses faster and typically they are younger than CTTS. Like CTTS HAeBes are surrounded by a disk and they actively accrete matter, but they are not expected to have an outer convective envelope, thus they should not generate magnetic fields and coronal activity. It is unclear if they can support magnetically funneled accretion. \citet{1994A&A...292..152Z} discovered X-ray emission from many HAeBes. More recent studies with the current generation of X-ray telescopes often identify the X-rays with a co-eval companion, i.e. a CTTS, but some HAeBes still seem to generate intrinsic X-ray emission \citep{2004ApJ...614..221S,2005ApJ...618..360H,2006A&A...457..223S,2009A&A...493.1109S}. There are two cases of HAeBes without any evidence of binarity and an existing X-ray grating spectrum: AB~Aur \citep{ABAur} and HD~163296 \citep{2005ApJ...628..811S,HD163296}.
\begin{figure}
\resizebox{\hsize}{!}{\includegraphics{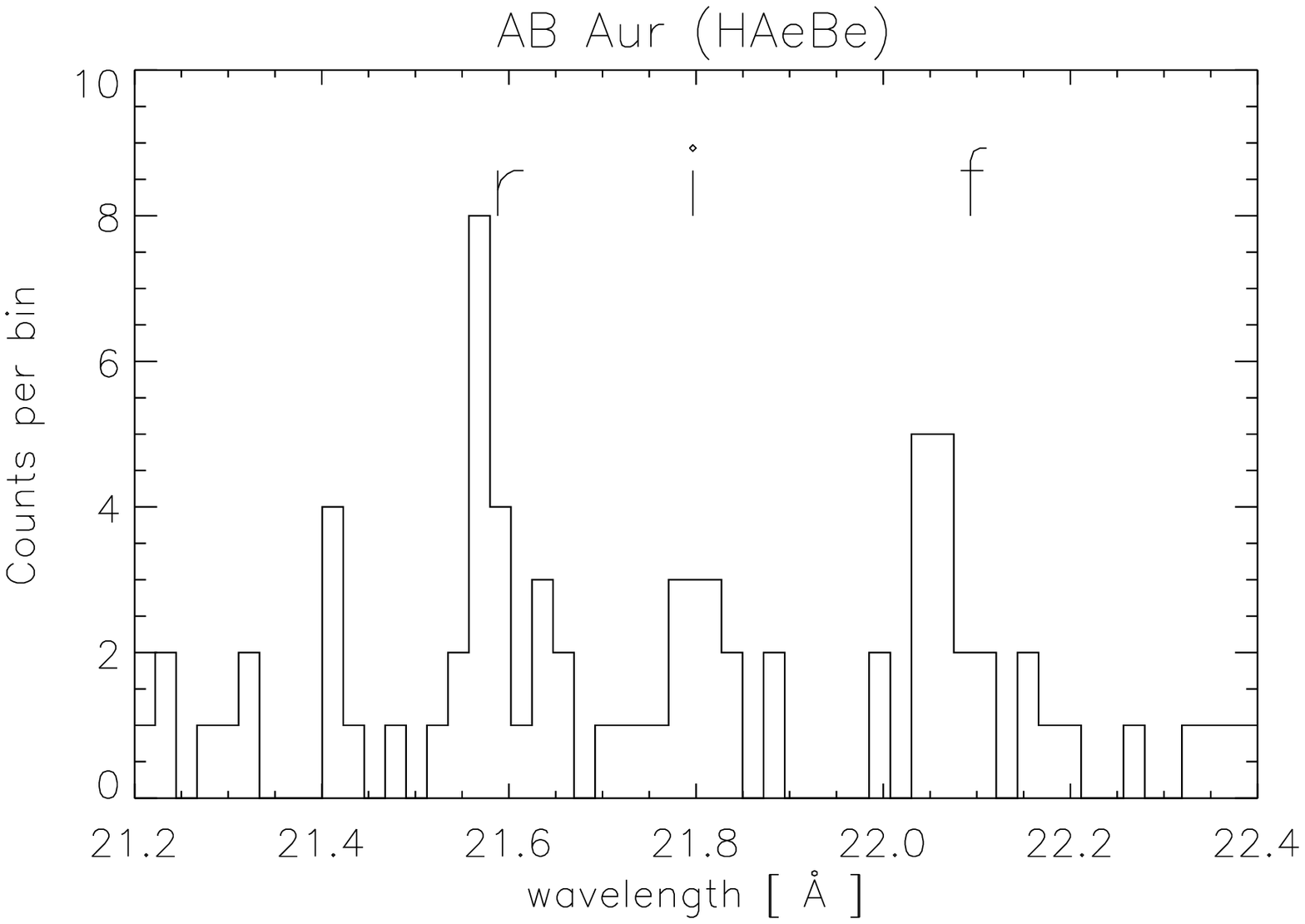}}
\resizebox{\hsize}{!}{\includegraphics{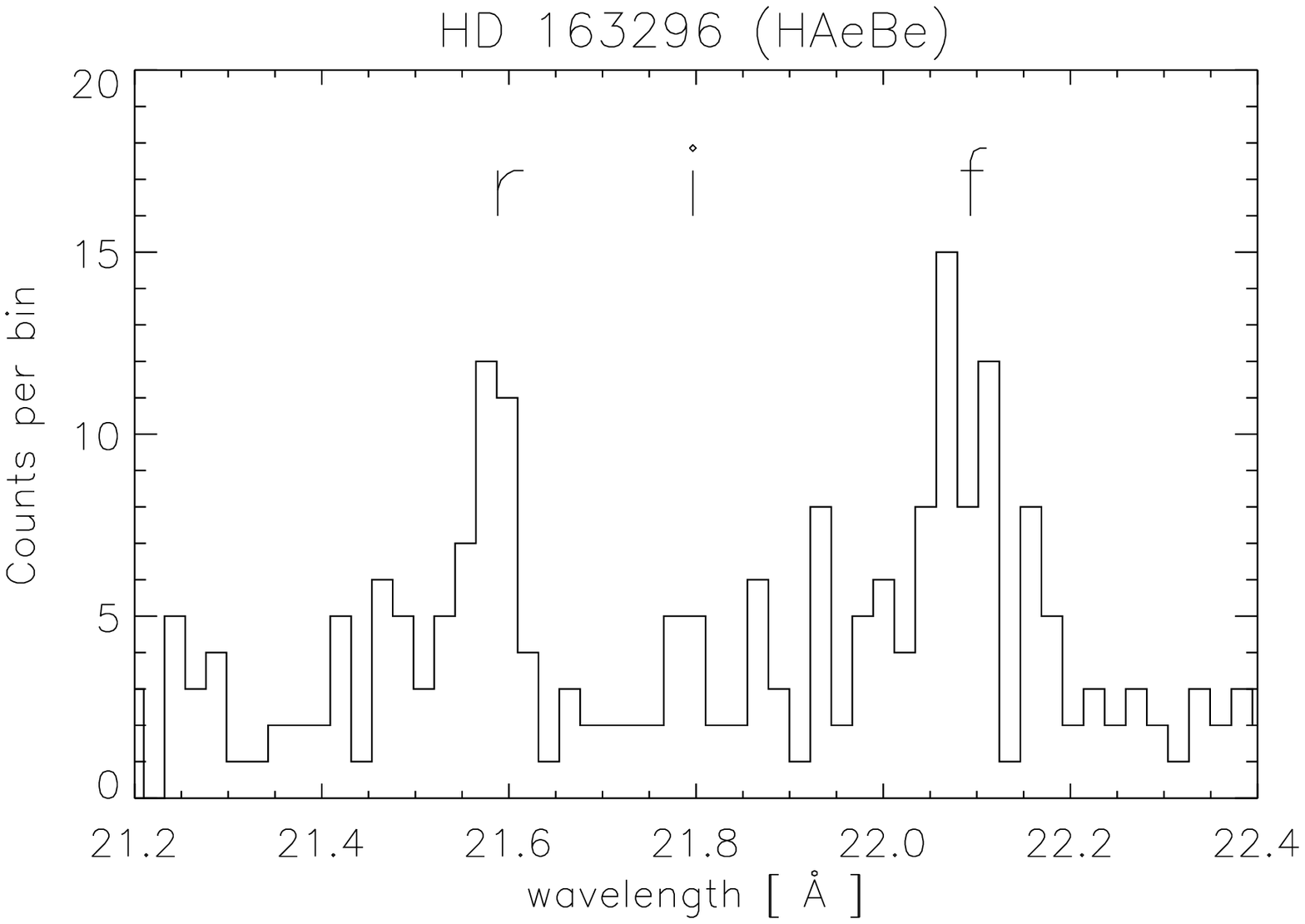}}
\caption{\ion{O}{vii} triplet for AB~Aur (top) and HD~163296 (bottom) observed with \emph{XMM-Newton}/RGS.  $r$, $i$ and $f$ line are labeled.}
\label{fig:haebetrip}
\end{figure}
Figure~\ref{fig:haebetrip} shows the \ion{O}{vii} triplets for those two HAeBes. The signal for AB~Aur is weak, because it falls on the edge of the detector. Both stars seem to have strong $f$ and weak $i$ lines, for HD~163296 $f/i>2.6$ on the 90\% confidence level \citep{HD163296}. This means that the emission originates in a region of low density and, given the strong UV field on the surface of A-type stars, this region is at an radius $R>1.7\ R_{*}$, i.e. at least $0.7\ R_*$ above the surface. Otherwise, the UV photons would shift emission from the $f$ to the $i$ line. Given the low absorbing column density it is unlikely that the hot plasma resides in a region closer to the surface, that is somehow shielded from the stellar radiation field. HD~163296 is known to drive a powerful jet, just as the CTTS DG~Tau, so a similar scenario, where the X-ray emission is caused by a shock in the jet, is possible. In fact, one knot in the jet of HD~163296 might itself be an X-ray source \citep{2005ApJ...628..811S}. In addition to this soft component, HAeBes also show surprisingly hot emission with fitted temperatures around 30~MK. This cannot be caused by accretion and is a clear signature of magnetic activity. HAeBes are not expected to drive solar-like dynamos, but they might operate turbulent dynamos in the atmosphere or retain the primordial magnetic field of the interstellar cloud. If the field lines are frozen-in the field strength increases by many orders of magnitude during the collapse to the proto-star. \citet{2004A&A...413..669G} observed a large flare in V892~Tau, a system of a HAeBe with a companion. The system is not resolved but they could tentatively identify the flare with the HAeBe star. This fits the picture of a magnetically heated corona.

The absence of high-density emission and the minimum distance between \ion{O}{vii} emission and the stellar surface shows that accretion does not contribute to the X-ray emission from HAeBes. Likely, the accretion process differs from CTTS, because the magnetic field is not strong enough to disrupt the disk at the co-rotation radius and to support magnetically funneled accretion. The mass may reach the star in the equatorial plane and form some kind of a boundary layer.

\section{Summary}
\label{sect:summary}
This article summarizes the results of high-resolution X-ray spectroscopy of young stars in the mass range of CTTS and HAeBes.

For two \emph{Chandra} observations I could deblend the \ion{Mg}{xi} triplet from the Ne Lyman series. It turns out that the Ne blend is weak, despite the enhanced Ne abundance.

At least three different processes contribute to the X-ray emission from CTTS, their importance varies between individual objects. First, there is a solar-like corona, second, the post-shock zone of the accretion shock cools radiating in X-rays and other wavelengths and, third, internal shocks in jets can heat matter to X-ray emitting temperatures. This is studied for the individual sources with the best spectra, but there is no reason why these results should not apply to CTTS as a class. The combined models explain the observed line ratios in the He-like triplets of \ion{O}{vii} and \ion{Ne}{ix} very well. 

The mass accretion rate required to power the X-ray emission is typically lower than values estimated from other wavelengths. This might be due to inhomogeneous spots, partial absorption or accretion streams, which impact at velocities significantly below the free-fall speed.

The mass flux in the X-ray component of the DG~Tau jets is also $10^{-3}$ of the total mass loss rate. This is fully compatible with optical observations, that show only the innermost jet component to be sufficiently fast (400 to 500~km~s$^{-1}$) to heat the gas in shocks to X-ray emitting temperatures. For an electron density $>10^5$~cm$^{-3}$ all dimensions of the shock cooling zone are only a few AU, so even in optical observations this cannot be resolved.

The circumstellar environment of some CTTS differs markedly from the interstellar gas-to-dust ratio. This can be explained, if the outflows of those CTTS are dust-depleted.

Last, the X-ray emission from HAeBes is likely intrinsic and not due to an unresolved companion. It can thus be established that HAeBes have a hot emission component similar to a solar-like corona. The line ratio in the \ion{O}{vii} He-like triplet rules out an accretion shock origin for the soft emission. Likely, a large fraction of the soft component is produced in the jet similar to DG~Tau.

\acknowledgements
I thank the Astronomische Gesellschaft for awarding me the newly established Promotionspreis in 2010. The work, which is summarized in this article, would not have been possible without the help and advice of my colleagues at the Hamburger Sternwarte and especially my thesis adviser Prof. J.~H.~M.~M. Schmitt. Jan-Uwe Ness helped a lot i nthe CORA fits to the Mg triplet. H.~M.~G. acknowledges financial support from Chandra grant GO6-7017X and from the Faculty of the European Space Astronomy Centre. This research made use of the Chandra Transmission Grating Catalog and archive (http://tgcat.mit.edu).

\bibliographystyle{../../../my_articles/aa-package/bibtex/aa} 
\bibliography{../../../my_articles/articles}

\end{document}